\documentstyle[sprocl,psfig]{article}

\begin{document}
\title{PHENOMENOLOGY OF ATMOSPHERIC NEUTRINOS
\footnote{Talk given by one of us (M.L.) in the 17th 
International Workshop on Weak Interactions and
Neutrinos, Cape Town, South Africa, January 1999}}
\author{{\bf Paolo Lipari} ~ and ~ {\bf Maurizio Lusignoli}}
\address{
Dipartimento di Fisica, Universit\`a ``La Sapienza", and \\ I.N.F.N.,
Sezione di Roma, Roma, Italy
}
\maketitle
\abstracts{
The relevance of the data concerning upward--going muons for the solution 
of the atmospheric neutrino problem is stressed.  
In particular, their inclusion in the analysis confirms the goodness of the 
neutrino oscillation hypothesis and allows to 
exclude some alternative, exotic explanations such as neutrino decay,  
flavour changing neutral currents, violations of the equivalence 
principle (at least in their simplest forms), and also to
discriminate, in principle,  between different neutrino oscillation models 
($\nu_\mu \leftrightarrow \nu_\tau$ versus $\nu_\mu \leftrightarrow \nu_s$), 
because of the difference in the  matter effects.
}

The measurements of the  fluxes of  atmospheric  neutrinos
by the  Super--Kamiokande (SK)  experiment  
\cite{SK-evidence,SK-data} show evidence
for the disappearance of muon  (anti)--neutrinos.
The same indication comes from the results of the MACRO
experiment~\cite{MACRO}.

While the simplest description  of the atmospheric  neutrino  data  
is given in terms of
$\nu_\mu \leftrightarrow \nu_\tau$ oscillations \cite{SK-evidence},
several other physical mechanisms  have  been  proposed in
the literature as viable explanations of the effect,
and in particular neutrino decay \cite{nu-decay},  
flavor  changing neutral  currents (FCNC) \cite{FCNC},
and violations of the equivalence  principle (VEP)
\cite {VEP} or, equivalently, of Lorentz
invariance  \cite{GHKLP}.
All these models have the common feature of `disappearing'  muon neutrinos,
however the   probability depends in different  ways  on the 
neutrino energy and  pathlength. 
For the contained or partially contained (sub--GeV and multi--GeV) 
events, the energy range of the parent neutrino is rather limited
(less than a few GeV) and therefore it is difficult to distinguish the 
different energy dependences from these data alone. A much wider energy 
region (median $E_\nu \sim 100$~GeV) can be studied looking at the 
upward--throughgoing muons, while the upward--stopping muons give 
independent information on the same energy region sampled by the 
multi--GeV (semi)--contained events. 
The three  `exotic'  models, at least in their simplest form,
are unable to fit at the same time the SK data for leptons 
generated inside the detector (sub-- and multi--GeV) and for up--going muons
generated in the rock near to it, and can therefore be excluded as the
main mechanism causing $\nu_\mu$ disappearance \cite{exotic}. 

In the fits to the 545 days SK data \cite{SK-data}, we introduced
a single parameter ($\alpha$) to allow for the uncertainty in normalization 
of the predictions in all the processes cosidered 
(including $\nu_e$--induced events). The systematic uncertainties
have been ignored in the definition of $\chi^2$ (that is therefore pessimistic). 
The flavor $\nu_\mu \leftrightarrow \nu_\tau$ oscillation model gives an 
excellent fit ($\chi^2$ = 33.3 for 32 d.o.f.) with maximal mixing, 
$\Delta m^2$ = $3.16\,10^{-3}$~eV$^2$ and $\alpha$ = 1.145.

The disappearance probability, that for flavor oscillations has the
well known form
\begin{equation}
P^{osc}_{\nu_\mu \to \nu_\tau}  = 
\sin^2 2 \theta ~\sin^2 \left [ {\Delta m^2 \over 4} \; {L \over E_\nu}
\right ],
\label{P-osc}
\end{equation}
has instead different forms for the exotic models considered:
\begin{eqnarray}
(\nu \ \ {\rm decay}) \ \ \ \ \ 
 P &=&\; 1 - \left \{ \sin^4 \theta + \cos^4 \theta \; \exp \left
(- {m_\nu \over \tau_\nu} {L\over E_\nu} \right ) \right \},
\label{P-decay2} \\
({\rm FCNC}) \ \ \ \ \ \ \ 
P &=&\;
 \frac{4\epsilon^2}{4\epsilon^2+\epsilon'^2}\;
~\sin^2 \left [ {G_F \over \sqrt{2}}\,  X_f \,\sqrt{4\epsilon^2+\epsilon'^2} 
 \right ],
\label{P-FCNC} \\
({\rm VEP}) \ \ \ \ \ \ \ \ 
P &=&\;
\sin^2(2 \theta_G) \;
\sin^2[ \delta |\phi| ~E_\nu \,L ].
\label{P-equiv}
\end{eqnarray}
In (\ref{P-decay2}), $P$ still depends on $L / E_\nu$, but with a different 
functional form; in (\ref{P-equiv}) the variable is $L \cdot E_\nu$; in
(\ref{P-FCNC}) $P$ is
independent on the neutrino energy $E_\nu$ and only 
depends on the column density, $X_f = \int_0^L dL' ~N_f(L') $,  
of the fermion target ($d$ quarks in our calculations) on which the 
muon (tau) neutrinos scatter    
nondiagonally ($\epsilon$) or with different strengths ($\epsilon'$). 

The SK data have the following features:
($a$) the e--like events are compatible with the no--oscillation prediction;
($b$) the sub--GeV $\mu$--like events are less than expected even for low zenith
angles, the suppression increasing with the angle;
($c$) the multi--GeV $\mu$--like events show no suppression for downgoing muons
and a suppression factor of about one half for upgoing (this is the strongest
signal for ``new physics");
($d$) the stopping upgoing muons are suppressed by $\sim 1/2$, except in the
bin nearest to horizontal;
($e$) the upward passing muons are less suppressed, and the shape of their
angular distribution is slightly deformed. 
All these features are well described by the flavor oscillation hypothesis, 
but the other models are unable to reproduce all of them.

\begin{figure} [hbt]
\centerline{\psfig{figure=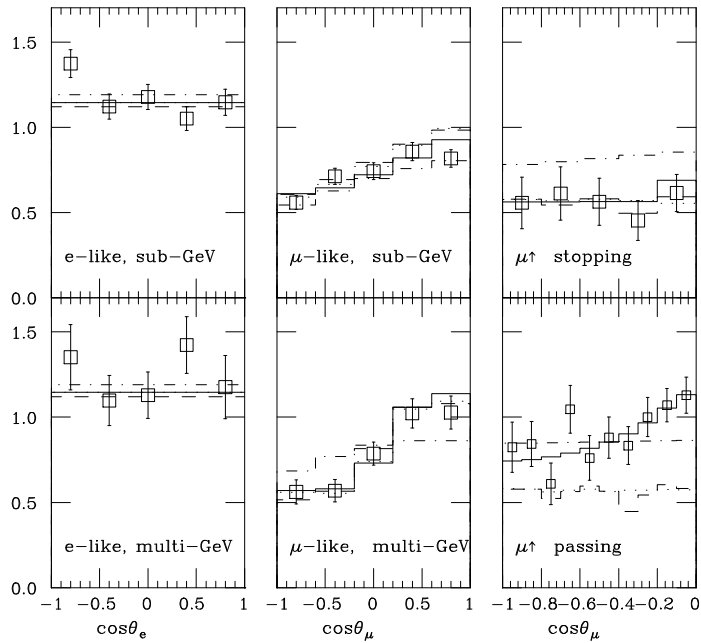,height=8cm}}
\caption {Ratio data/MonteCarlo for the $\mu$--like events in SK.
The histograms give the best fit predictions for oscillations (solid),
$\nu$ decay (dot-dash), FCNC (dashes) and VEP model (dots).
\label{fig_data}
}
\end{figure}

This is shown in fig.~\ref{fig_data}, where the ratio data/MonteCarlo 
is plotted for
the SK events, and the best--fit predictions are given 
for the normal oscillations (full lines), the neutrino decay (dot-dashed),
the FCNC (dashed) and the VEP model (dotted).
Numerically, the $\chi^2$ values are 33.3, 82, 149, 143, respectively, for
32 degrees of freedom. Parameter values are: $\tau_\nu/m_\nu = 18840$ Km/GeV,
$\cos^2 \theta =  0.84$ and $\alpha =1.19$ for neutrino decay;
$\epsilon = 1.4$ and $\alpha = 1.12$ (we assumed maximal mixing, 
 $\epsilon' = 0$) for FCNC; 
$\delta |\phi| = 4.5 \cdot 10^{-4}$ Km$^{-1}$GeV$^{-1}$, $\theta_G = \pi/4$ and
$\alpha = 1.145$ for the VEP model. It is to be noted that ignoring the
upward--going muon data and fitting only (semi)--contained events we obtain 
best--fits (with somewhat different parameter values) having $\chi^2$ equal to 
25, 39, 35, 38, respectively, for 18 d.o.f.: 
allowing for the systematic errors and differences in normalization the exotic 
models may become acceptable in this case. The larger neutrino energy range
covered by the upward muon events is essential to rule out the different energy
dependences of FCNC and VEP models. A more detailed discussion has been given 
elsewhere~\cite{exotic}.

Even if oscillations give the favoured solution of the atmospheric 
neutrino anomaly, and the Chooz experiment~\cite{Chooz} excludes a dominant
oscillation involving electron neutrinos, one has still the open possibilities 
of $\nu_\mu$ oscillating mainly in $\nu_\tau$ or in a fourth, sterile neutrino. 
These can be distinguished looking at neutral current 
events~\cite{Vissani-Smirnov} and also by a careful study highlighting the
different behaviours due to the matter effects~\cite{sterile,LL98}.
In this respect, the higher energy upward muon data could be essential.

In fact, the relevance of the matter effects depends on the neutrino energy,
and in particular on the quantity $\zeta = (2 \,E_\nu \,V_{\mu s}) / 
\Delta m^2$ (where $V_{\mu s} = \mp\sqrt{2}\,G_F\, N_n / 2$ 
is the difference in effective potentials, for $\nu$ or $\bar{\nu}$). 
For $|\zeta| \ll 1$ the matter effects  are  negligible.
For $|\zeta| \gg 1$ the matter effects  are dominant and 
the oscillations are strongly supressed: the
effective mixing   $\sin^2 2\,\theta_m$ decreases like 
$\zeta^{-2}$ and the oscillation length levels off to a value 
$\ell_m^\infty = 2\,\pi\,/\,|V_{\mu s}|
 \simeq 1.3 \,(5\,{\rm g}\,{\rm cm}^{-3}/\rho)\;10^4$~km, independent from 
$E_\nu$ and $\Delta m^2$, and remarkably
close to the earth's diameter.
In the region $|\zeta| \sim  1$,  that corresponds to 
a neutrino energy
$E_\nu   \sim 5.2~ {{\rm GeV}}$ 
 $\left ( |\Delta m^2|/ 10^{-3}~\rm {eV}^2 \right )$ 
$\left (  5\,{\rm g}\,{\rm cm}^{-3} / \rho \right )$,
one has the most complex behavior.
The analyses of the sub--GeV  and multi--GeV  data in terms of 
neutrino oscillations suggest for $|\Delta m^2|$ a value 
in the range $10^{-3} \div 10^{-2}$~eV$^2$, therefore 
matter  effects  are essentially negligible in 
the  sub--GeV region, they can  be  relevant 
in the multi--GeV  region, but only 
if $|\Delta m^2|$ is close to the lower end of the above range, 
and are always significant for upward  throughgoing muons.

\begin{figure} [hbt]
\centerline{\psfig{figure=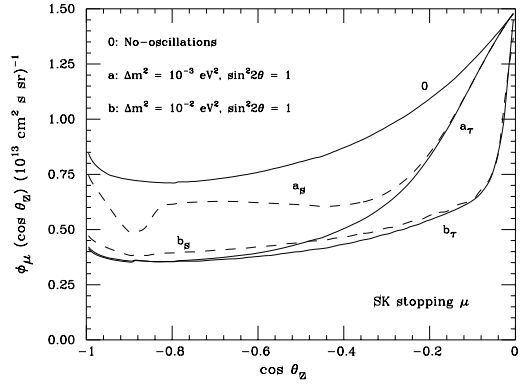,height=5cm}}
\caption{Predictions of the rate  and angular  distribution of
stopping muons  in SK,  in the absence  (and presence) of oscillations.
The subscript $\tau$  (s) indicates
$\nu_\mu$--$\nu_\tau$
($\nu_\mu$--$\nu_s$) mixing.
\label{fig_stop}
}
\end{figure}

\begin{figure} [hbt]
\centerline{\psfig{figure=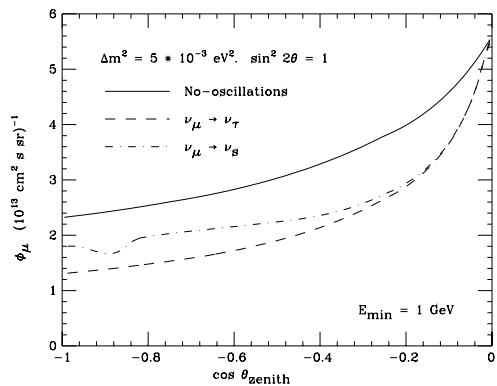,height=5cm}}
\caption{Upward--going muon flux as a function of zenith angle (with
$E_{\rm min} = 1$~GeV), in the absence of oscillations
(solid line), and  for  maximal mixing
and  $\Delta m^2 = 5 \cdot 10^{-3}$~eV$^2$
in the cases  of $\nu_\mu \to \nu_\tau$ (dashes) and
$\nu_\mu \to \nu_s$ (dot--dashes).
\label{fig_upmu0} }
\end{figure}

As an example, we show in fig.~\ref{fig_stop} the prediction 
for upward muons stopping in SK, where one can easily tell the difference for
$\Delta m^2$ = $10^{-3}$~eV$^2$, but not so for $10^{-2}$~eV$^2$.
In fig.~\ref{fig_upmu0} we report the prediction for the flux of 
upward going muons with
an energy larger than 1 GeV (appropriate to the MACRO experiment), showing
the different deformations in shape that are expected for maximal mixing 
and $\Delta m^2$ = $5 \cdot 10^{-3}$~eV$^2$. In both figures
one can see a dip for $\cos \theta \simeq -0.9$, that has been widely
discussed and differently interpreted\cite{reson} in recent times. 
In a more detailed work \cite{LL98} we have also reported results of
calculations on the (smaller) matter effects for contained events.

In conclusion, we stress the importance of the upward muon data to gain
a complete and thorough understanding of the atmospheric neutrino
anomaly.

\newpage

\end{document}